\documentclass{iau}

\usepackage{amsmath}
\usepackage{graphicx}
\usepackage{multirow}

\newcommand{\aap}{    {\it A\&A}}

\newcommand{\apj}{    {\it ApJ}}
\newcommand{\apjl}{   {\it ApJL}}
 
\newcommand{\apjs}{    {\it ApJS}}

\newcommand{\grl}{    {\it Geophys. Res. Lett.}}

\newcommand{\pasp}{   {\it Pub. Astron. Soc. Pac.}}

\newcommand{\solphys}{{\it Solar Phys.}}
 
\newcommand{\ssr}{    {\it Space Sci. Revs}} 

\newcommand{\kms}{   {km s$^{-1}$}}
\chardef\us=`\_

\begin{document}

\lefttitle{Devi et al.}
\righttitle{EUV Wave and Coronal Seismology}

\jnlPage{1}{7}
\jnlDoiYr{2024}
\doival{10.1017/xxxxx}

\aopheadtitle{Proceedings IAU Symposium 388}
\editors{N. Gopalswamy,  O. Malandraki, A. Vidotto \&  W. Manchester, eds.}

\title{EUV Wave and Coronal Seismology } %\LaTeX\ Guidelines for~authors}
%\title{\color{red}{EUV Wave and magnetic field seismology of the Solar Corona}} %\LaTeX\ Guidelines for~authors}

\author{Pooja Devi$^1$, Ramesh Chandra$^1$, Arun Kumar Awasthi$^2$, Brigitte Schmieder$^{3,4}$, Reetika Joshi$^{5,6}$}

%\author[addressref={aff1},corref,email={setiapooja.ps@gmail.com}]%{\inits{P.}\fnm{Pooja}~\lnm{Devi}\orcid{0000-0003-0713-0329}}

%\address[id=aff1]{Department of Physics, DSB Campus, Kumaun University, Nainital 263 001, India}

\affiliation{$^1$Department of Physics, DSB Campus, Kumaun University, Nainital 263001, India\\
$^2$Space Research Centre, Polish Academy of Sciences, Bartycka 18A, 00-716 Warsaw, Poland\\
$^3$LIRA, Observatoire de Paris, CNRS, Universit\'e de Paris, 5 place Janssen, 92290 Meudon, France\\
$^4$Centre for mathematical Plasma Astrophysics, Dept. of Mathematics, KU Leuven, 3001 Leuven, Belgium\\
$^5$Rosseland Centre for Solar Physics, University of Oslo, P.O. Box 1029 Blindern, N-0315 Oslo, Norway\\
$^6$Institute of Theoretical Astrophysics, University of Oslo, P.O. Box 1029 Blindern, N-0315 Oslo, Norway\\
Email:setiapooja.ps@gmail.com}

\begin{abstract}
%We present the investigation of the Extreme-Ultraviolet (EUV) wave associated with a flare that occurred on 28 October 2021 and associated coronal loop oscillation and type II radio burst. 
{We present an investigation of the Extreme-Ultraviolet (EUV) wave linked to the flare that occurred on 28 October 2021, along with the associated coronal loop oscillation and type II radio burst.}
The EUV wave was observed by multi-viewpoint with Solar Dynamics Observatory and Solar Terrestrial Relations Observatory -- A. The associated coronal mass ejection (CME) was observed by Large Angle and Spectrometric Coronagraph (LASCO) as well by COR1 coronagraph. From the multi-view observation, we found that the EUV wave is propagated ahead of the connected CME. The coronal magnetic field measurement was performed by the coronal loop oscillations as well by the associated m-type II radio burst observations. We found the magnetic field strength values computed by both methods are consistence and are in the range of $\approx$ 5 to 10 G.
\end{abstract}


\maketitle

\section{Introduction}
%{\bf General introduction}
\label{intro}
During the campaign of H$\alpha$ observation in 1960, \cite{moreton1960} found the large scale moving disturbances on the solar chromosphere. Later these disturbances were named as ``Moreton waves''. Their reported speed is in the range of 500 to 2000 \kms\ \citep[see reviews by][]{Warmuth2015, ChenPF2016}. The generation of Moreton waves is due to the fast mode magnetohydrodynamic (MHD) shock sweeping into the chromosphere. 
\cite{Thompson1998} observed similar propagating features in 195 \AA\ wavelength with Extreme-ultraviolet Imaging Telescope \citep[EIT;][]{Delaboudiniere1995} onboard Solar and Heliospheric Observatory \citep[SOHO;][]{Domingo1995} and named them as ``EIT waves''. Because of their clear visibility in Extreme-ultraviolet (EUV) wavelengths, they are popularly known as ``EUV waves'' and their speed ranges from a few hundreds to multiple of thousand \kms.
%Since it is clearly and easily observable in EUV wavelengths 

Initially, these waves were observed with low-cadence data of SOHO and Solar Terrestrial Relations Observatory (STEREO) satellites \citep{thompson1999, Zhukov2004, Zhukov2009} and were proposed to be the fast-mode MHD waves. However, after analyzing the high spatio-temporal data of Solar Dynamics Observatory (SDO), two components (slow, speed is as low as 10 \kms and fast (speed can be up to multiple of 1000 \kms) as well as stationary fronts of this phenomena were reported \citep{Chen2011, Asai2012, Chandra2016, Chandra2018, Chandra2021}. The observed features can be explained by the hybrid model proposed by \cite{Chen2002, Chen2005}. 
According to this model, a piston-driven shock forms along the expanding flux rope during its eruption, sweeping the whole solar surface. The legs of the shock produce fast-mode EUV waves. On the other hand, the slower (or nonwave) component, forms due to successive opening or stretching of the magnetic field lines straddling over the erupting flux rope.

%{\bf Relation with CMEs and type II:}\\
When the coronal mass ejection (CME) propagates out of the solar surface, it produces a shock wave and it is believed that this shock wave could be responsible for the EUV wave disturbances \citep{Biesecker2002}. There are studies, where people try to look into the association between the EUV waves and CMEs and it is found that these two physical phenomena are closely associated \citep{ Chen2009, Dai2010, Schmieder2013, Nitta2014, Muhr2014}. Another evidence of the shock wave is the type II radio bursts observed in coronal heights and can be associated with the EUV waves. 

%{\bf Loop Oscillations:}
When the fast component of the EUV wave encounters the surrounding magnetic structures, these structures can become unstable. These magnetic structures can be filament or coronal loops. As a result of this instability, these structures can oscillate and be used for the coronal seismology \citep{Warmuth2005, Ballai2007, Guo2015, Luna2017, Fulara2019, Shen2019, Devi2022}. 

%{\bf Structure of the article:}
%In this article, we present an EUV wave of 28 October 2021 associated with a large GOES X1.0 eruptive solar flare, nearby loop oscillations, and type II radio burst. In section \ref{observations}, the observational data sources along with the results are presented. The summary of the article is given in section 3.
In this article, we present an EUV wave from 28 October 2021, associated with a large GOES X1.0 eruptive solar flare, nearby loop oscillations, and a type II radio burst. Section~\ref{observations} covers the observational data sources and the results. The summary of the article is provided in Section~\ref{summary}.

\section{Observations}
\label{observations}
%\textbf{About data source and event description}
The EUV wave event of 28 October 2021 was observed by several space based telecopes and its counterpart i.e., Moreton wave, by ground based instruments. We analyse the data of Atmospheric Imaging Assembly \citep[AIA;][]{Lemen2012} onboard Solar Dynamics Observatory \citep[SDO;][]{Pesnell2012} and Global Oscillation Network Group \citep[GONG;][]{Harvey2011} from Earth view and by Extreme-ultraviolet Imager (EUVI) and coronagraphs (COR1) onboard ahead spacecraft of Solar Terrestrial Relations Observatory \citep[STEREO--A;][]{Kaiser2008} from a different view angle. 
{The radio emission due to this event were observed by e-Callisto spectrographs at Birr Castle, Ireland (BIR) and also by ORFEES \citep{Hamini2021} at Nan\c{c}ay Radio Observatory. 
}
%Here, we use the data of AIA in 171 and 193 \AA\ and STEREO-A EUVI in 195 \AA\ wavelengths along with its COR1 coronagraph.
%
The EUV wave event is accompanied by one of the major GOES X1.0 class two ribbon flare and filament eruption, which produced a ground level enhancement (GLE). The origin of the activities was from NOAA AR 12887, which was located at S26W07 on 28 October 2021. 
The studies have been done on different activities produced by this event which include: different features of the EUV and Moreton wave \citep{Hou2022, Devi2022}, solar energetic particles \citep{Kouloumvakos2024} and related radio bursts \citep{Klein2022}, and GLE \citep{Mavromichalaki2022, Papaioannou2022, Mishev2024}. Along with the observations, data-driven simulations of the event are performed by \cite{Guo2023}.

\subsection{Morphology of EUV Wave and associated CME}

\begin{figure}[!t]
    \centering
    \includegraphics[width=10cm]{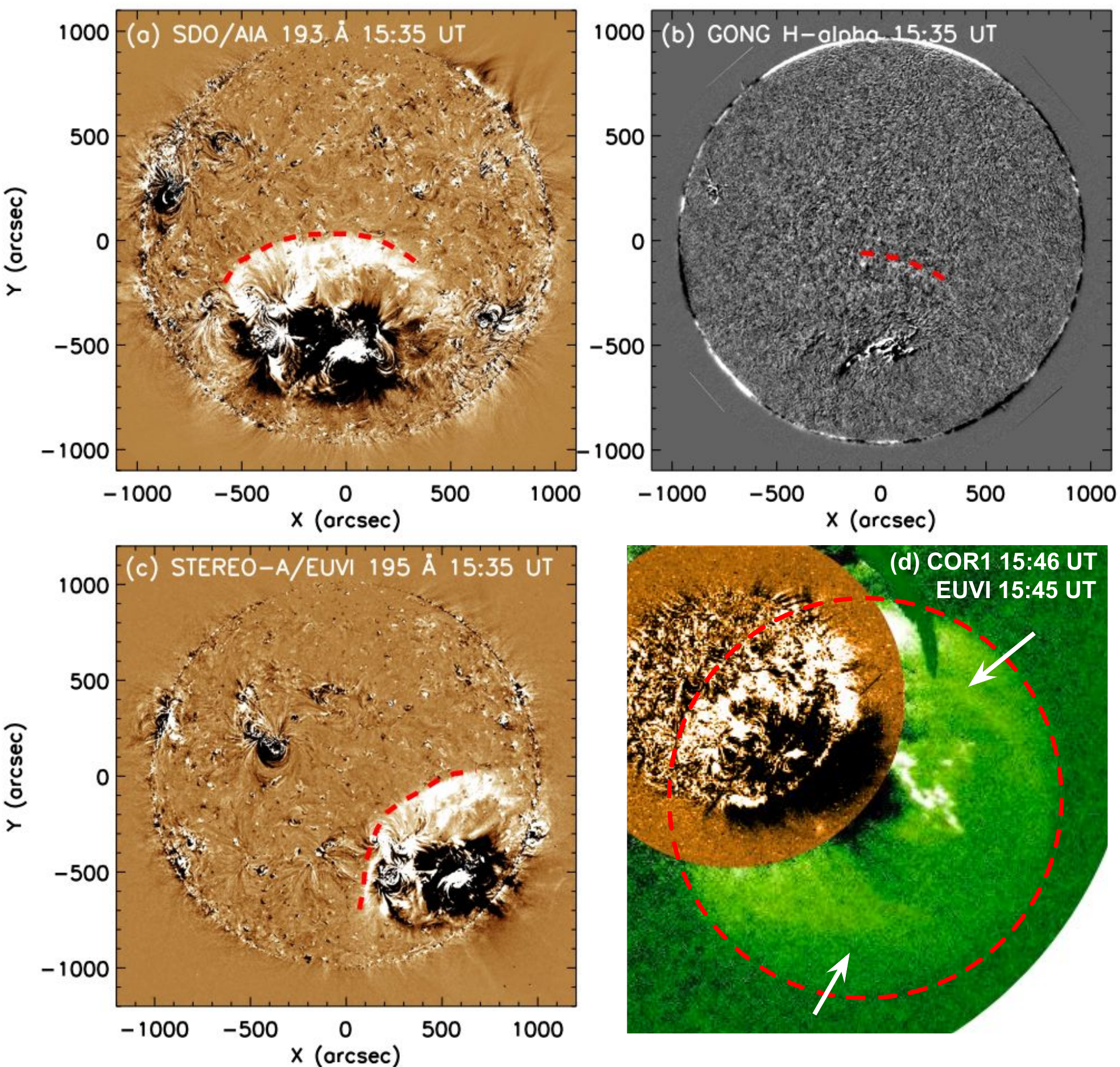}
    \caption{Panel (a, c): Snapshots of the EUV waves observed in AIA 193 \AA, and STEREO-A 195 \AA. Panel b: H$\alpha$ Observation of GONG, where Moreton wave is visible. panel (d): Overlay of CME observed by COR1 with the STEREO-A data. The red dashed curves in panels (a -- c) and the circle in panel (d) are the fronts of the wave. The white arrows in panel (d) show the CME leading edge. \citep[Panels (a) and (d) are adapted and recreated from][]{Devi2022}.}
    \label{wave_cme}
\end{figure}

The EUV wave appeared on the solar disk $\sim$ 15:25 UT in a circular shape. The propagating EUV wave was promptly visible in the northward and fainter in the southward direction. Figure \ref{wave_cme} displays base difference snapshots of the wave in different wavelengths. 
%The front of EUV and the Moreton wave is shown by the red dashed curve in panel (a, c) and (b), respectively. 
The kinematics of the wave were done by \cite{Devi2022} and \cite{Hou2022} using the time-distance technique. 
Two components (i.e., fast-mode and nonwave) of the EUV wave were observed. The speed of the nonwave component was found to be $\approx$ 280 \kms\ and fast-mode ranges from $\approx$ 550 to 720 \kms.
The speed of the Moreton wave observed in H$\alpha$ was found in the range of $\approx$ 310 to 540 \kms.

The EUV wave was associated with a halo CME of linear speed 1519 \kms and an acceleration of -61 m s$^{-2}$ as reported by the LASCO CME catalog. 
The CME was also observed by the STEREO-A COR1 coronagraph. 
In the STEREO-A data, the event is observed close to the western limb, which provides us to an excellent opportunity to compare spatially these two phenomena.
For this comparison, the image of EUVI is overlaid on the COR1 image as shown in Figure \ref{wave_cme}(d). The internal yellow part is the EUVI 195 \AA\ image at 15:45 UT and the outer green part is the COR1 image at 15:46 UT. Assuming the spherical structure of the EUV wave, we fitted a circle at the outer edge of the EUV wave (shown by a red dashed circle) as suggested by \cite{Gopalswamy2013}. The alignment between the EUVI and COR1 images evidenced that the EUV wave was traveling ahead of the CME. This result is confirmed by the data-driven simulation of \cite{Guo2023} (see their Figure 10).

\subsection{Loop Oscillations}
\begin{figure}
    \centering
    \includegraphics[width=10cm]{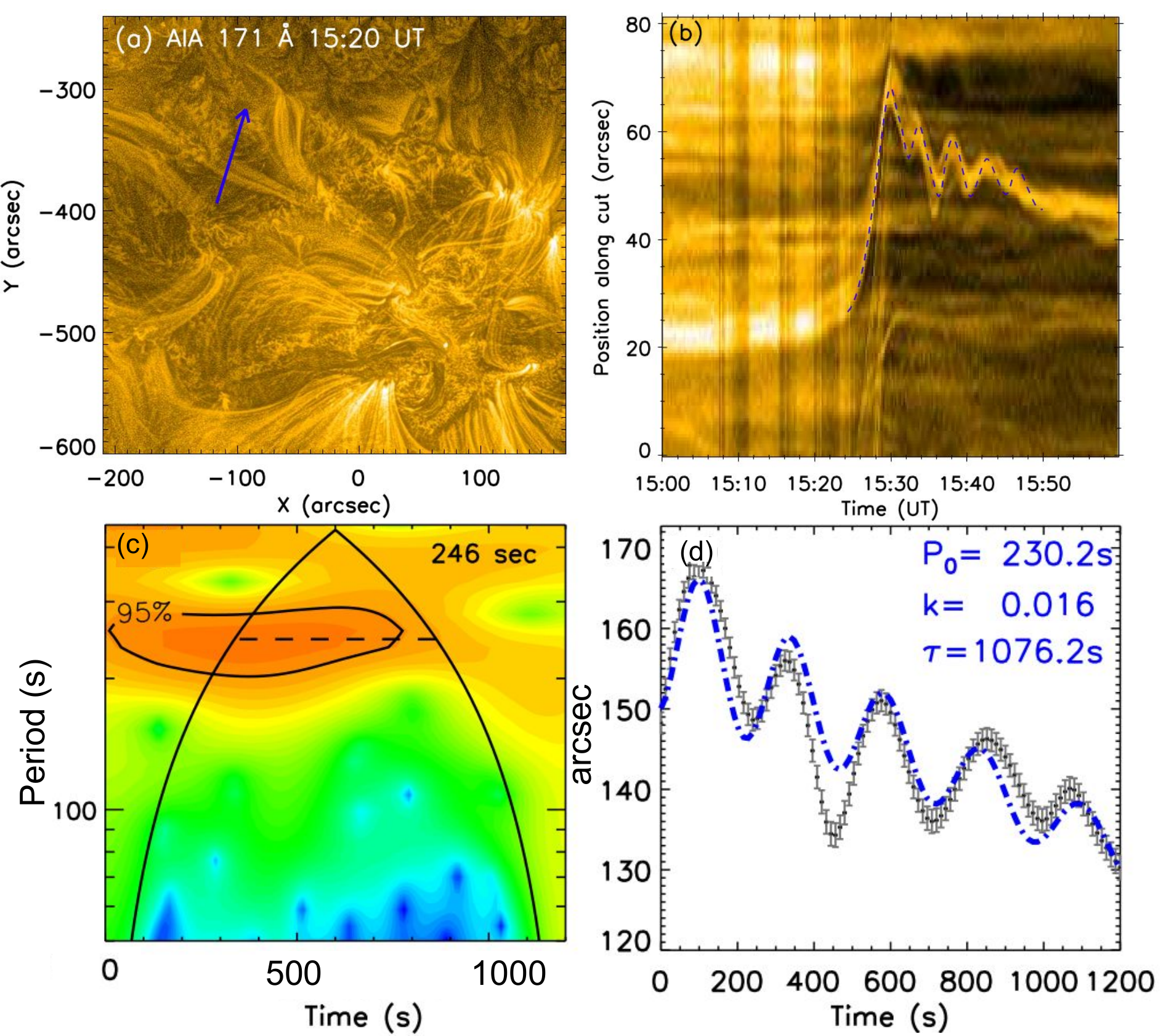}
    \caption{AIA 171 \AA\ image with the clearly observed oscillating loop system is depicted in (a) panel. The direction of oscillation is in the blue arrow direction. The time-slice plot showing the oscillations is shown in (b) panel. The wavelet analysis along with the temporal evolution of oscillation (overlaid by dashed lines) is displayed in panel (c) and (d) panel of the figure. (Adapted from \cite{Devi2022}) }
    \label{loop}
\end{figure}

During the propagation of the EUV wave in the north direction, it encountered the neighboring coronal loops and triggered them to oscillate. The oscillations of two loop systems have been studied in detail by \cite{Devi2022}. They computed the periods of loop oscillations using the wavelet method and by fitting a damped sinusoidal function given in their equation 1. The periods from both these methods are found to be consistent with each other and are in the range from 230 to 280 sec. 
They also calculated the densities inside (n$_{in}$) and outside (n$_{out}$) the oscillating loops using the Differential Emission Measure \citep[DEM;][]{Cheung2015} technique which are found in the range 1.6$\times$10$^{8}$ -- 2.1$\times$10$^8$ cm$^{-3}$ and 1$\times$10$^8$ -- 1.8$\times$10$^8$ cm$^{-3}$, respectively. Oscillations in one of the loop systems along with its periodic analysis are presented in Figure \ref{loop}. Further, \cite{Devi2022} calculated the coronal magnetic field strength (B) using the following equation \citep{Roberts1984},
\begin{equation}
    B=\frac{L}{P}\sqrt{8\pi\mu m_p n_{in}\Big(1+\frac{n_{out}}{n_{in}} \Big)}
\end{equation}

where, $\mu$ is the average molecular weight for coronal abundances, and $m_p$ is mass of proton. Putting all the above values along with the length (L) of oscillating loop (115 -- 175 Mm), the B found to vary from $\approx$ 5.7 to 8.8 G. 
%where $a_0$, $a_1$, and $a_2$ are amplitude related constants, $\tau$, $\phi$, and $P_0$ are damping time, initial phase, and period of oscillation, respectively. The constant $k$ represents a linear variation of the period in time. The period of oscillation of loops is found to be consistent with both the methods described above, which is close to 4 minutes.

%The period of oscillations can be used in deriving the coronal magnetic field strength (B) by using the formula proposed by \cite{Roberts1984}, which is given by the following equation,

%\begin{equation}
%    B=\frac{L}{P}\sqrt{8\pi\mu m_p n_{in}\Big(1+\frac{n_{out}}{n_{in}} \Big)}
%\end{equation}

%where, L and P are the length and period of the oscillating loop, $\mu$ is the average molecular weight for coronal abundances, and $m_p$ is mass of proton. $n_{in}$ and $n_{out}$ are the densities inside and outside of the loop. \cite{Devi2022} calculated these densities using Differential Emission Measure \citep[DEM;][]{Cheung2015} and found 1.6 $\times$ 10$^8$ and 1 $\times$ 10$^8$ cm$^{-3}$ inside and outside the loop, respectively. After the value of L which is 175 Mm and all other calculated values, we get the value of B in the range 8.2 to 8.8 G. Considering the oscillations other loop systems as well (which are not shown here), the B ranges from $\approx$ 5.7 to 8.8 G (please visit \cite{Devi2022} for more details). 

\subsection{Type II Radio Burst}
\begin{figure}
\centering
\includegraphics[width=10cm]{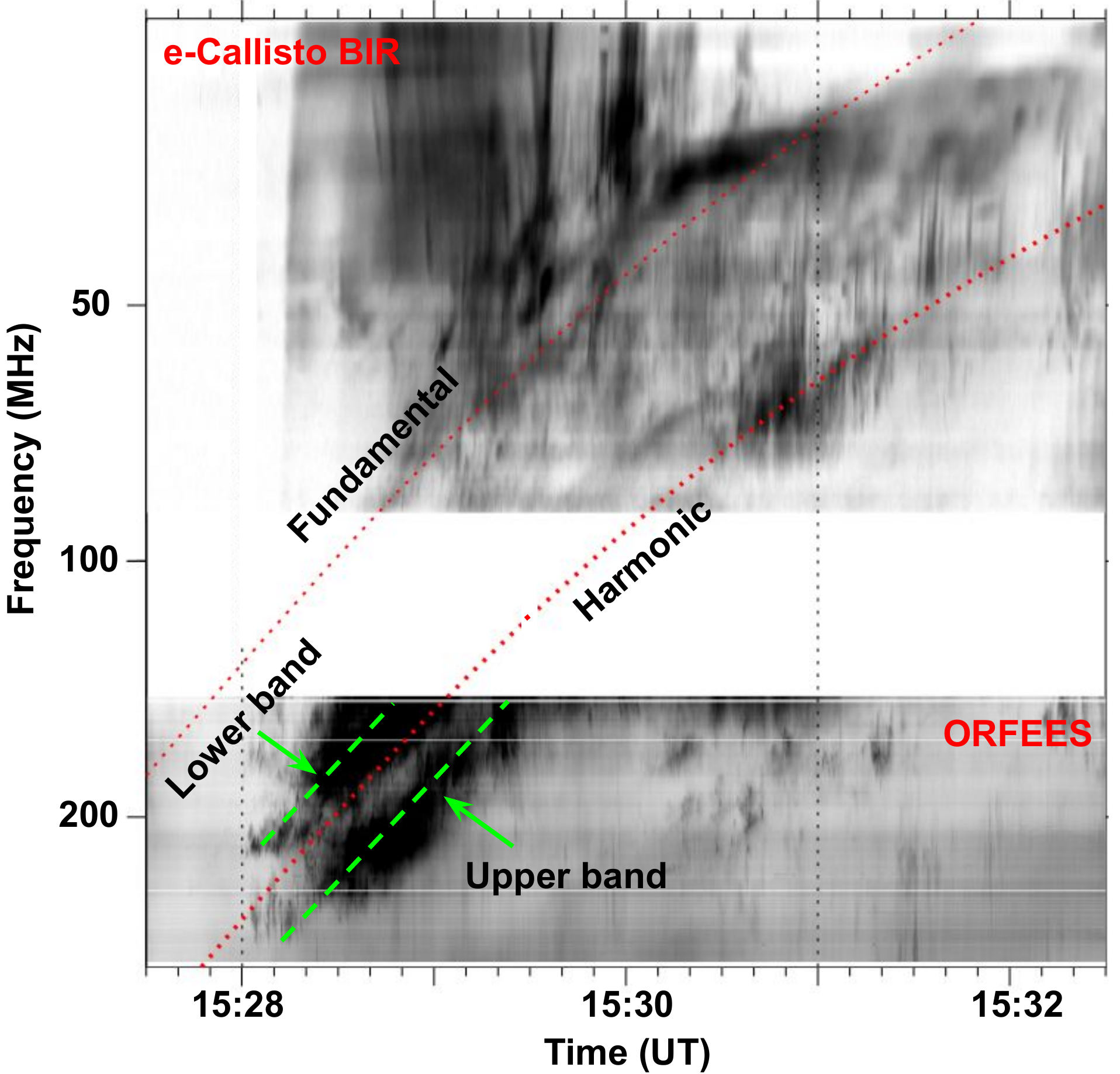}
\caption{Type II radio bursts in metric range by combining the dynamic spectra from BIR (e-Callisto at Birr Castle, Ireland) and ORFEES. The fundamental and harmonic emissions are shown by red dotted curves. The splitted upper and lower bands of harmonic emission are denoted with green arrows. (Adapted and reformed from \cite{Klein2022}) }
\label{typeII}
\end{figure}

The event was also associated with type II radio bursts in metric (m) and decameteric-hectometeric frequency range. 
The detailed study about the observed type II burst was investigated by \cite{Klein2022} (see their figures 2, 5, and 12). They computed the speed of m-type II burst in the range 1900 -- 2600 \kms. The observed metric type II radio burst is presented in Figure \ref{typeII}. The type II radio burst shows the harmonic and fundamental structures. The splitting in the harmonic structure is observed in ORFEES dynamic spectra. Such splitting in the harmonic band is due to the consequence of the plasma emission of the upstream and downstream shock regions and it was initially interpreted by \cite{Smerd1974}. The frequency of observed upstream (f$_u$) and downstream (f$_l$) are 173 and 130 MHz, respectively. Using the parameters of harmonic band splitting of m-type II bursts \cite{Klein2022} calculated the Alfv\'en Mach number (M$_A$) which is 1.6. 
%They calculated the Mach number (M$_A$) using the parameters of type II radio burst in metric wavelength by using the formula given by
%\begin{equation}
%M_A = \sqrt{\frac{X(X+5)}{2(4-X)} },
%\end{equation}
%where, $X = ({f_u}/{f_l})^2$, $f_u$ and $f_l$ are the upper and lower band frequencies. Putting the values of $f_u$ = $\approx$ 103 Mm and $f_l$ $\approx$ 77 Mm, the M$_A$ is found to be 1.6.
Now, using the parameters computed by them, we have calculated B by the following formula \citep{Vasanth2014},
\begin{equation}
\label{mag_field_typeII}
B = 5.1 \times 10^{-5} \times v_a \times f_l,
\end{equation}

where the Alfv\'en speed ($v_a$={$v_s$}/{M$_A$}) found to vary from 1187 to 1625 \kms. 
%where $v_a$ is the Alfv\'en speed (equal to {$v_s$}/{M$_A$}) found to vary from 1187 to 1625 \kms. 
Inserting the values of $v_a$ and $f_l$ in equation \ref{mag_field_typeII}, the computed B is found in the range $\approx$ 7 -- 10 G, which is consistent with B  calculated using the loop oscillation parameters (Section \ref{loop}).

\section{Summary}
\label{summary}
The EUV wave of 28 October 2021 showed both the fast-mode and nonwave components which can be explained by the hybrid model proposed by \cite{Chen2002}. Our analysis revealed that the EUV wave travels in front of the CME leading edge implying that the EUV wave is driven by the CME and not by the flare pressure pulse \citep{Vourlidas2009, Kwon2014, Kwon2017, Guo2023}. An important characteristic of this EUV wave is its association with the neighbouring loop oscillations and the type II radio bursts. 
We have computed the coronal magnetic field using the loops oscillation parameters which are 5.7 -- 8.8 G and by the m-type II radio observations harmonic band splitting, the computed value is about 7 -- 10 G. The computed B from both these methods are consistent with each other and in the range of previous studies \citep{Nakariakov2001, Aschwanden2011, Guo2015, Su2018, Zhang2022}.

%\bibliographystyle{apj}
%\bibliography{references}

\end{document}